\documentclass{PoS}

\usepackage{url}

\newcommand{\tcat}{2FLGC}
\newcommand{\lat}{\emph{Fermi}-LAT}
\newcommand{\tf}{$T_{05}$}
\newcommand{\tn}{$T_{90}$}

\title{High-energy emission from GRBs: \\ 10 years with {\it Fermi}-LAT}

\ShortTitle{High-energy emission from GRBs}

\author{\speaker{E. Bissaldi}$^{1,2}$, M. Axelsson$^{3,4}$, N. Omodei$^{5}$, and G. Vianello$^{5}$, $\;\;\;\;\;\;\;\;\;\;\;\;\;\;\;\;\;\;\;\;\;\;\;\;\;\;\;\;\;\;\;\;\;\;\;\;\;\;\;\;\;\;\;\;\;$ 
        on behalf of the Fermi-LAT Collaboration \\
        $^{1}$Dipartimento Interateneo di Fisica, Politecnico di Bari, Via G. Amendola 126, 70125 Bari, Italy\\
        $^{2}$Istituto Nazionale di Fisica Nucleare (INFN) - Sezione di Bari, Via E. Orabona 4, 70125 Bari, Italy\\
        $^{3}$Department of Physics and Oskar Klein Center for Cosmoparticle Physics, Stockholm University, SE-106 91 Stockholm, Sweden \\
        $^{4}$Department of Physics, KTH Royal Institute of Technology, AlbaNova, SE-106 91 Stockholm, Sweden \\
        $^{5}$W.W. Hansen Experimental Physics Laboratory, Kavli Institute for Particle Astrophysics and Cosmology, Department of Physics and SLAC National Accelerator Laboratory, Stanford University, Stanford, CA 94305, USA; \\
        E-mail: \email{elisabetta.bissaldi@ba.infn.it}}

\abstract{In 2018, the Fermi mission celebrated its first decade of operation. In this time, the Large Area Telescope (LAT) has been very successful in detecting the high-energy emission (>100 MeV) from Gamma-Ray Bursts (GRBs). The analysis of particularly remarkable events - such as GRB 080916C, GRB 090510 and GRB 130427A - has been presented in dedicated publications. Here we present the results of a new systematic search for high-energy emission from the full sample of GRBs detected in 10 years by the Fermi Gamma-Ray Burst Monitor, as well as Swift, AGILE, Integral and IPN bursts, featuring a detection efficiency more than 50\%\ better than previous works, and returning 186 detections during 10 years of LAT observations. This milestone marks a vast improvement from the 35 events contained in the first LAT GRB catalog (covering the first 3 years of Fermi operations). We assess the characteristics of the GRB population at high energy with unprecedented sensitivity, covering aspects such as temporal properties, energetics and spectral index of the high-energy emission. Finally, we show how the LAT observations can be used to inform theory, in particular the prospects for very high-energy emission.}

\FullConference{36th International Cosmic Ray Conference -ICRC2019-\\
		July 24th - August 1st, 2019\\
		Madison, WI, U.S.A.}
\begin{document}
\section{Introduction}
Gamma-ray bursts (GRBs) are the most luminous astrophysical events observed in the Universe. They are brief flashes of high-energy radiation emitted by an ultra-relativistic collimated outflow which is thought to originate from a stellar-mass black hole formed by the merging of binary systems (short GRBs) or the explosions or massive stars (long GRBs). Observations by the \emph{Fermi} Gamma-Ray Space Telescope in the ten years since it was placed into orbit on 2008 June 11 have given the opportunity to study the broadband properties of GRBs over an unprecedented energy range.  Its two scientific instruments, the Large Area Telescope (LAT) \cite{2009ApJ...697.1071A}, and the Gamma-Ray Burst Monitor (GBM) \cite{2009ApJ...702..791M}, provide the combined capability of probing emission from GRBs over seven decades in energy. These ground-breaking observations have helped to characterize the highest energy emission from these events, furthered our understanding of the emission processes associated with GRBs, helped place constraints on the Lorentz invariance of the speed of light, the gamma-ray opacity of the Universe, and motivated revisions in our basic understanding of collision-less relativistic shock physics. Here we summarize the main results from the first 10-year period of observations with \lat, covering the period from August 2008 to August 2018. A full account is presented in the second catalog of GRBs detected by \lat\ (\tcat) \cite{2019ApJ...878...52A}.
\section{The second {\it Fermi}-LAT GRB catalog}
The \tcat\ presents the results of a search for high-energy counterparts of GRBs that triggered space instruments and have an available public localization. After accounting for bursts that triggered more than one instrument, we had a total of 3044 independent GRBs. We used the localizations provided by the GBM, unless a better localization was provided by one of the instruments on-board the {\it Neil Gehrels Swift Observatory} \cite{2004ApJ...611.1005G}, namely the Burst and Transient Alert Telescipe (BAT), the X--Ray Telescope (XRT), or the UV--Optical Telescope (UVOT), or by the Interplanetary Network (IPN)\footnote{\url{http://www.ssl.berkeley.edu/ipn3/}}.

For each trigger, a dedicated search was run to look for a high-energy counterpart in standard LAT data above 100 MeV. The search was first performed using a dedicated detection algorithm, searching five different time windows from 10\,s to 10\,ks. All triggers which passed the criteria for detection in any time window were manually inspected, to rule out other sources, such as Earth limb contamination or flaring AGN. Only detections retained after this inspection were passed to the analysis pipeline. Here, analyses were performed to determine a number of key properties for each GRB, such as onset, duration, and spectral parameters. The onset ($T_{\,\rm LAT,0}$) is calculated as the arrival time of the first photon with at least 90\% probability of being associated with the GRB, and the duration ($T_{\, \rm LAT,1}$) is the period from the first to the last such photon. All these quantities were calculated between 100 MeV and 100 GeV.

A separate detection analysis was run to see if the GRB was found in the LAT Low Energy (LLE) data, which covers the range from 30 MeV to 1 GeV. If so, temporal properties also in this range were determined. Onset and duration are defined using the ``standard'' quantities of \tf\ and \tn, where \tf\ is the time at which 5\% of the total emission as been received, and \tn\ is the period from \tf\ until 95\% of the emission has been detected.

In total, 186 GRBs (17 short and 169 long) were detected by \lat\ in the 10 years analysed. Of these, 169 were seen above 100 MeV and 91 were found in LLE data; 17 of which were only detected below 100 MeV. This gives the largest sample of high-energy GRB detections so far.
\begin{figure}[t!]
\centering
\begin{tabular}{ccc}
\includegraphics[height=0.28\columnwidth]{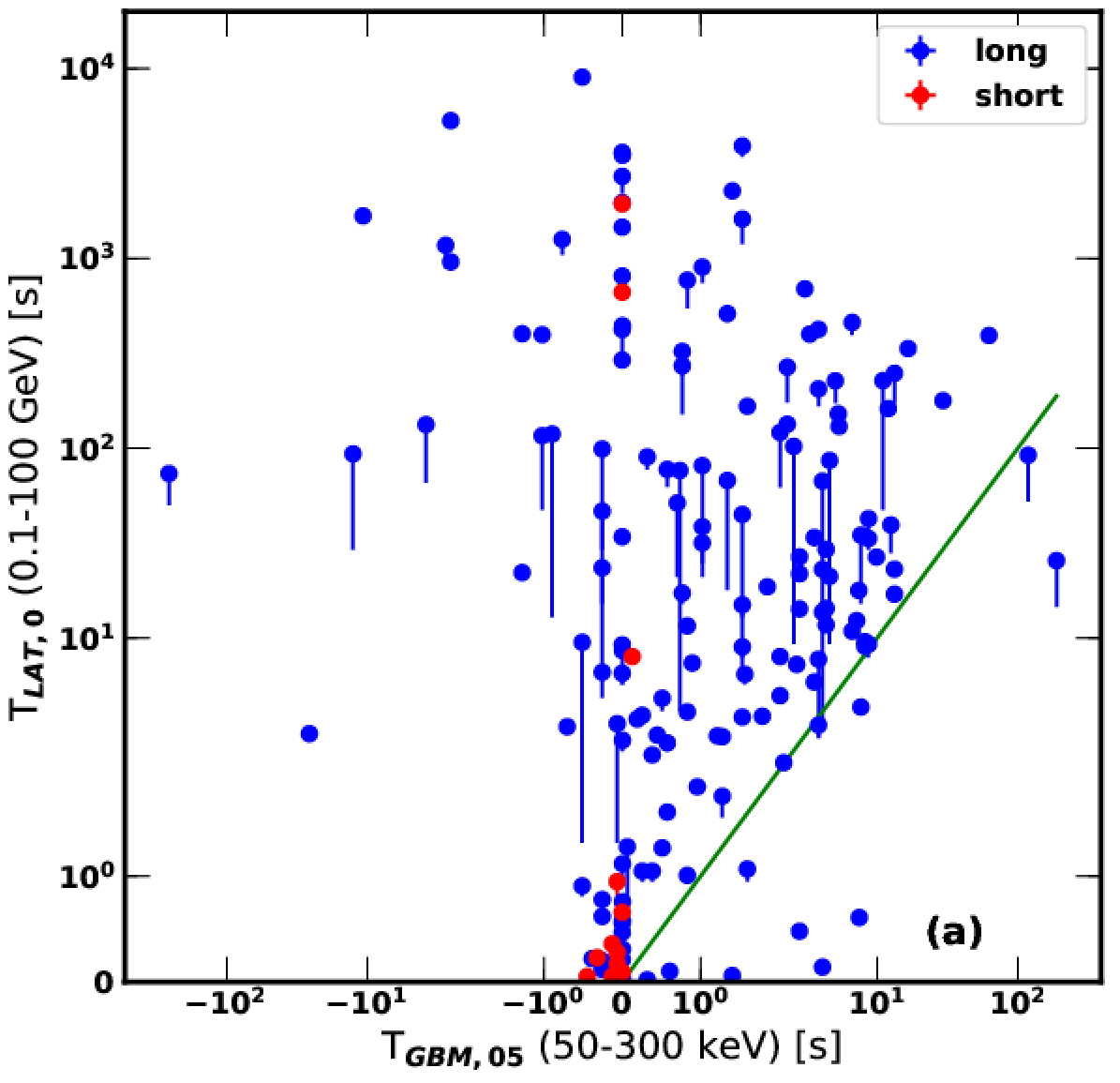} &
\includegraphics[height=0.28\columnwidth]{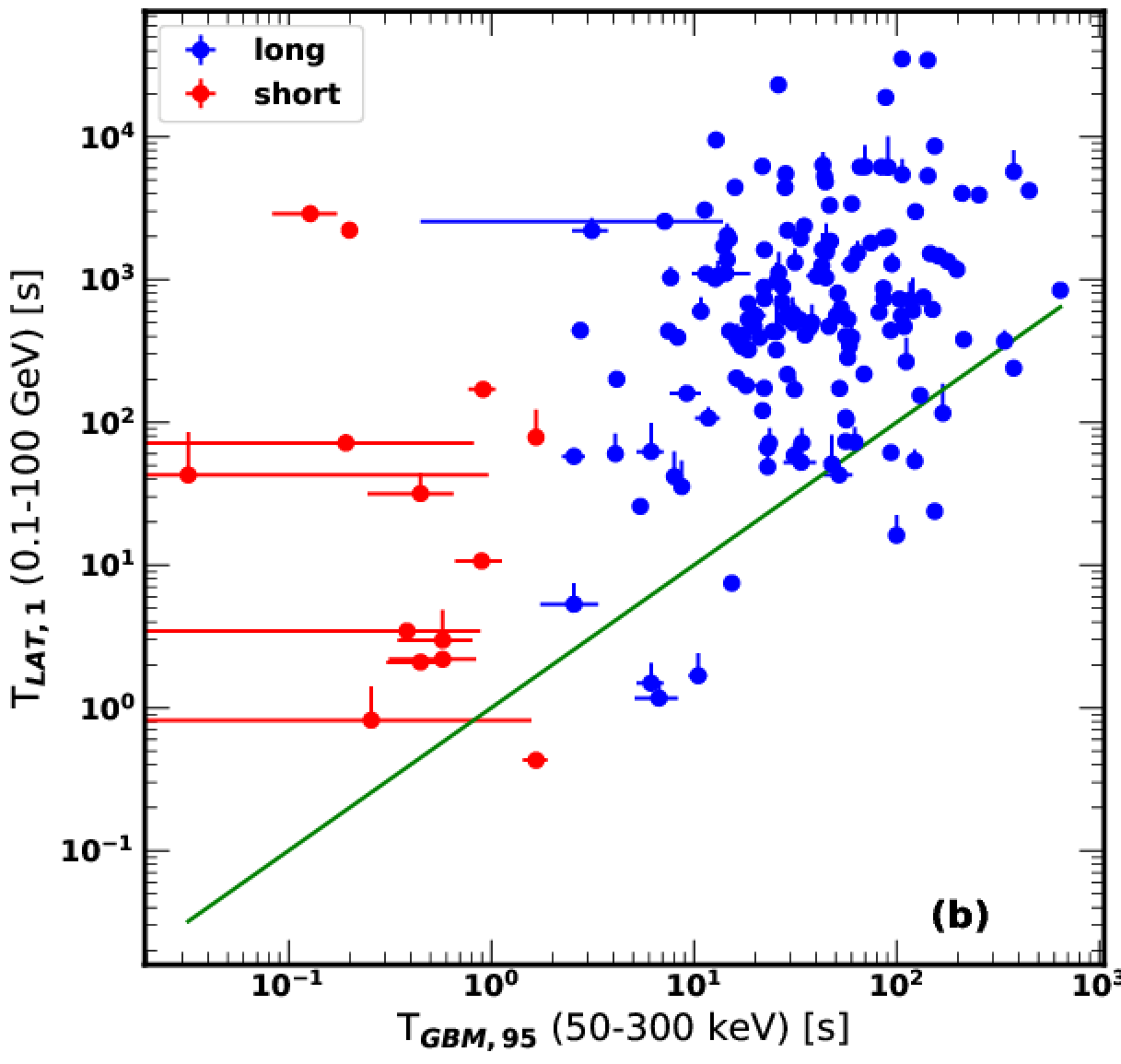} &
\includegraphics[height=0.28\columnwidth]{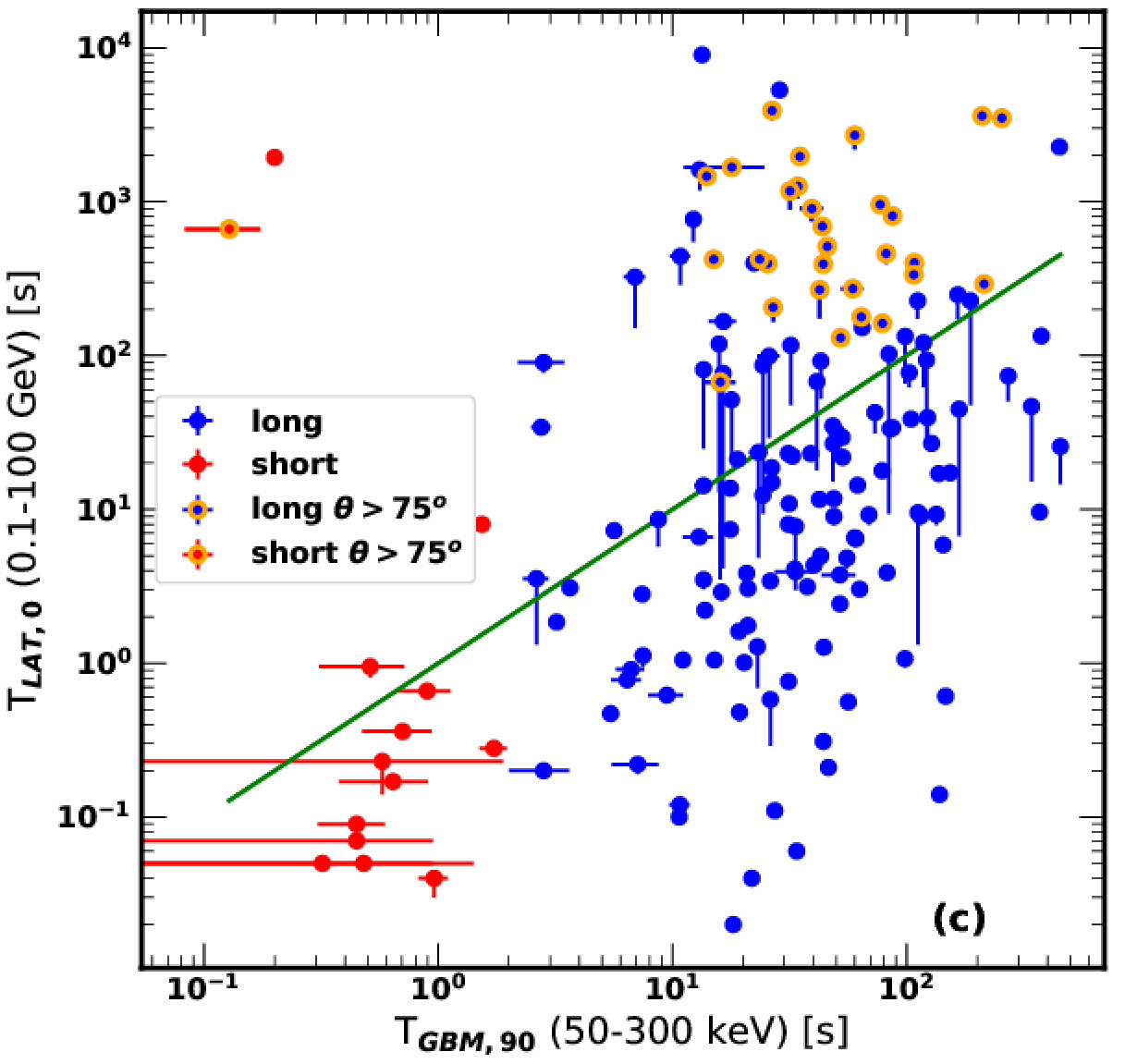}
\end{tabular}
\caption{
Onset (a) and duration (b) calculated in the 100 MeV--100 GeV energy range vs. the same quantities calculated in the 50--300 keV energy range. Panel (c) shows the onset times in the 0.1--100 GeV vs. the durations in the 50--300 keV energy range. The solid line denotes where values are equal. Blue and red circles represent long and short GRBs, respectively. In panel (c), we additionally mark the GRBs which were outside FoV at trigger time with a thick orange contour. Figure from \cite{2019ApJ...878...52A}.}
\label{fig_temporal_LAT}
\end{figure}
\section{Temporal properties}
The first \lat\ GRB catalog \cite{2013ApJS..209...11A}, which covered only 3 years of data and included 35 GRBs, showed that emission above 100 MeV is typically delayed compared to that at lower energies, but also lasts longer. Figure~\ref{fig_temporal_LAT} shows the same comparison of onset and duration at high (0.1--100 GeV) and low (50-300 keV) energies for the current study. The latter quantities are taken from the {\emph{Fermi}}-GBM Online Burst Catalog\footnote{\url{https://heasarc.gsfc.nasa.gov/W3Browse/fermi/fermigbrst.html}}.

As can be seen in the figure, our results confirm and strongly support the claim that when high--energy emission is observed in GRBs, this emission is delayed and lasts longer compared to that in the low--energy band. This also fits with the scenario where the two energy bands probe different physical processes (e.g., internal emission from the jet and external shocks). However, panel (c), showing the onset time of the high--energy emission versus the burst duration in the 50--300 keV range, adds further information. It is clear that in the majority of events, the high energy emission starts already during the period of the prompt emission phase (as defined by the low-energy emission). This means that whatever process is behind the high-energy emission, it has to be active very early on.

\begin{figure}[t!]
\centering
\begin{tabular}{cc}
\includegraphics[height=0.4\columnwidth]{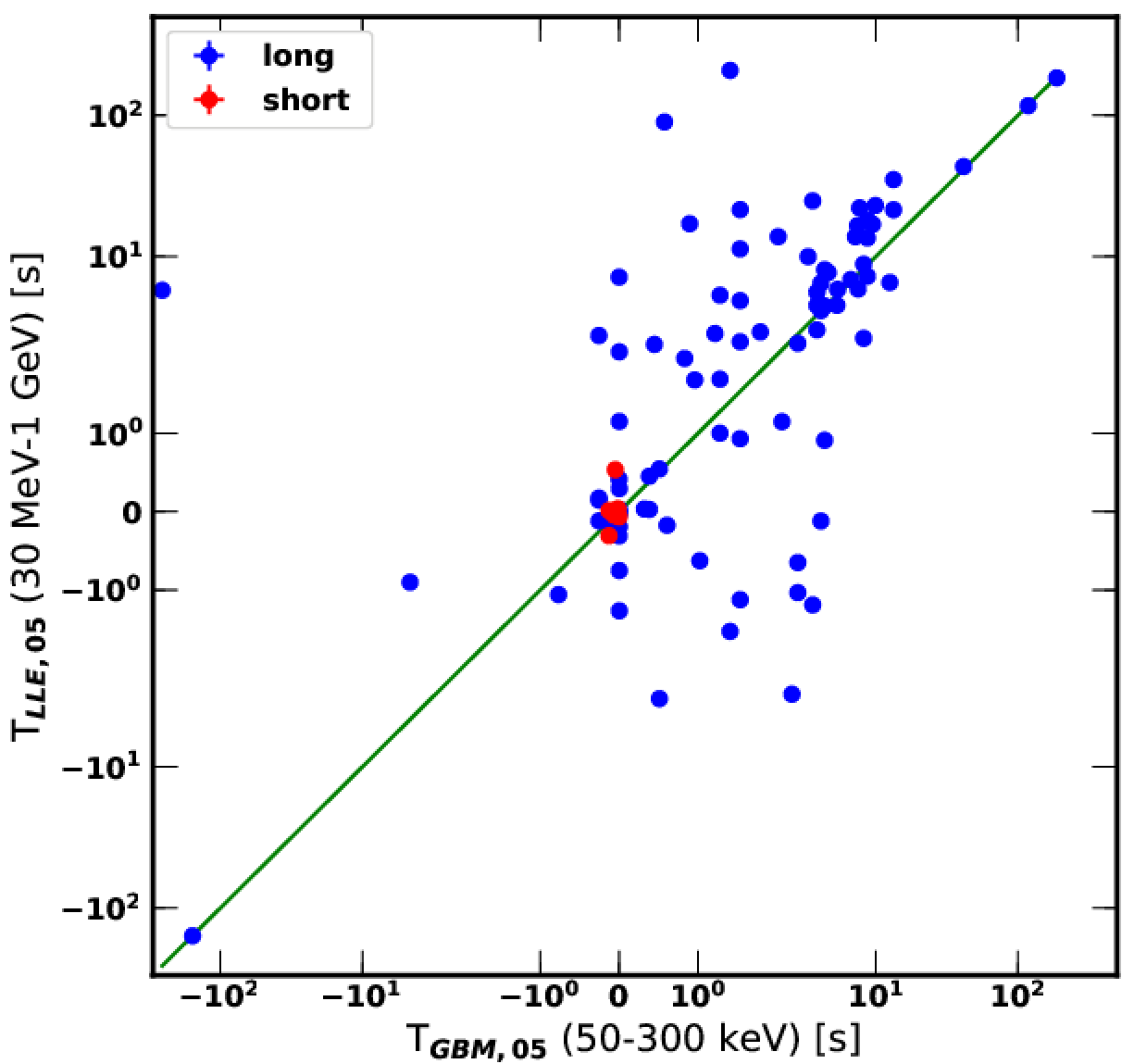} & 
\includegraphics[height=0.4\columnwidth]{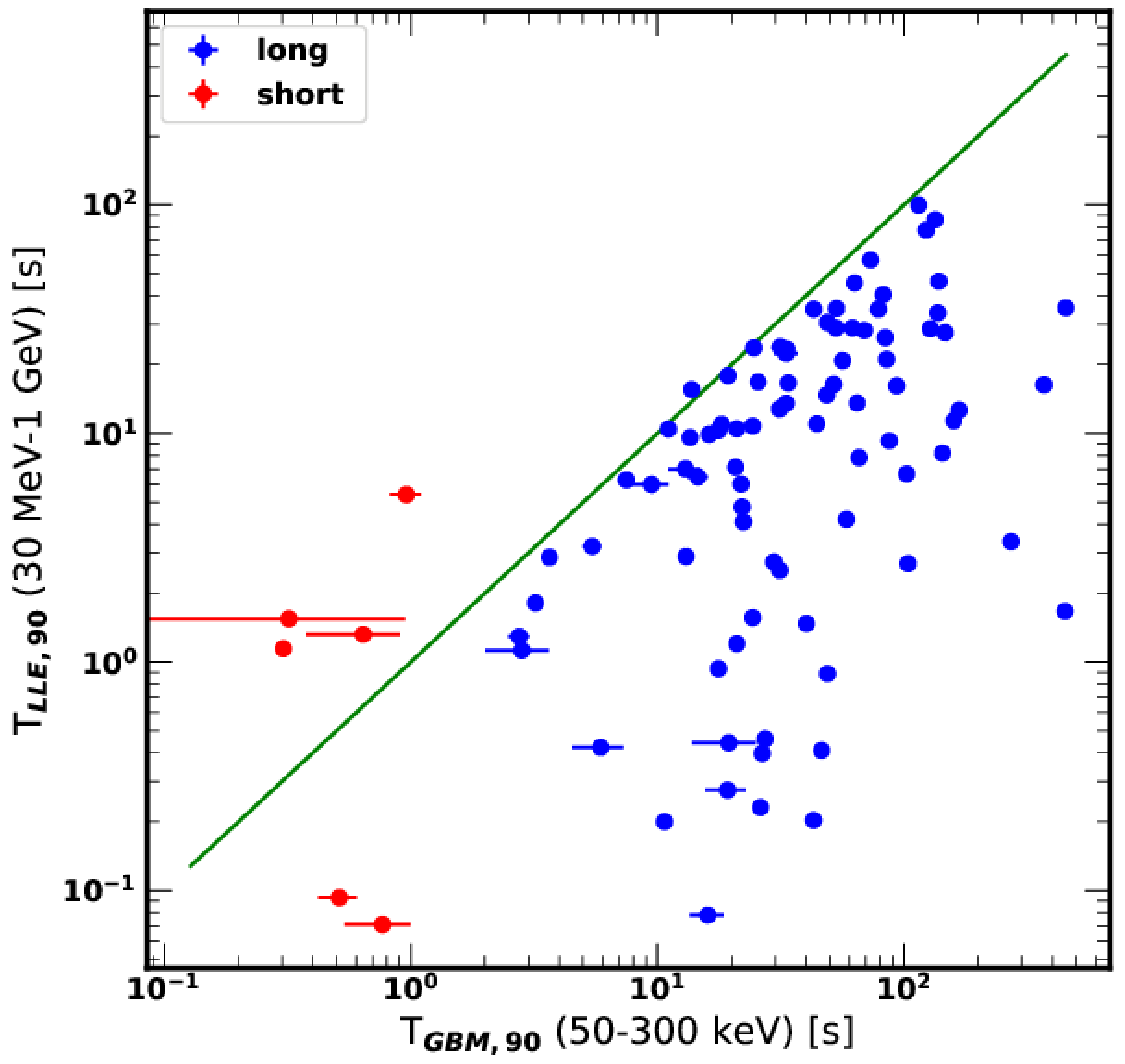}
\end{tabular}
\caption{Onset times (left panel) and durations (right panel) calculated using LLE data in the 30 MeV--1 GeV energy range vs. the corresponding quantities calculated using GBM data in the 50--300 keV energy range. The solid green line denotes where value are equal. Blue and red circles represent long and short GRBs, respectively. Figure from \cite{2019ApJ...878...52A}.}
\label{fig_temporal_LLE}
\end{figure}

Figure~\ref{fig_temporal_LLE} shows the comparison between the onset and duration in the 30 MeV--1 GeV range covered by LLE and that at 50--300 keV. The onset time is here very similar in both energy ranges, while the duration is significantly shorter in the LLE range. Since the LLE data is dominated by photons in the lower part of the range, the similarities indicate that the component seen at 50--300 keV is driving the emission up to at least 30 MeV. The shorter duration in the LLE range is consistent with the hard-to-soft evolution often seen at lower energies. This behavior has previously been reported by {\it e.g.} \cite{2002ApJ...579..386N} using BATSE data, as well as for several LAT-observed GRBs \cite{2012ApJ...757L..31A, PoS(IFS2017)065, 2018ApJ...864..163V}, and we can here confirm that it applies to the vast majority of GRBs (in particular long GRBs).
\subsection{Temporal decay}
As noted above, many GRBs show long-lasting emission at high energies. The large sample of GRBs in our sample enables us to measure the temporal decay index of this emission. In total, 88 GRBs had sufficient data to provide a reliable decay index, assuming a power-law decay. The mean value of the index is $0.99 \pm 0.04$ with a standard deviation of $0.80 \pm 0.07$. The stable decay index is an indication that the high-energy emission could be related to the same component that is responsible for the afterglow emission in the X-band. We also note that some GRBs are better fit with a broken power-law, or show signs of variability in the lightcurve. These are again in some ways similar to the flares and plateaus seen in X-rays.
\section{Spectral index and energetics}
\begin{figure}[t!]
\centering
\begin{tabular}{ccc}
\includegraphics[height=0.28\textwidth]{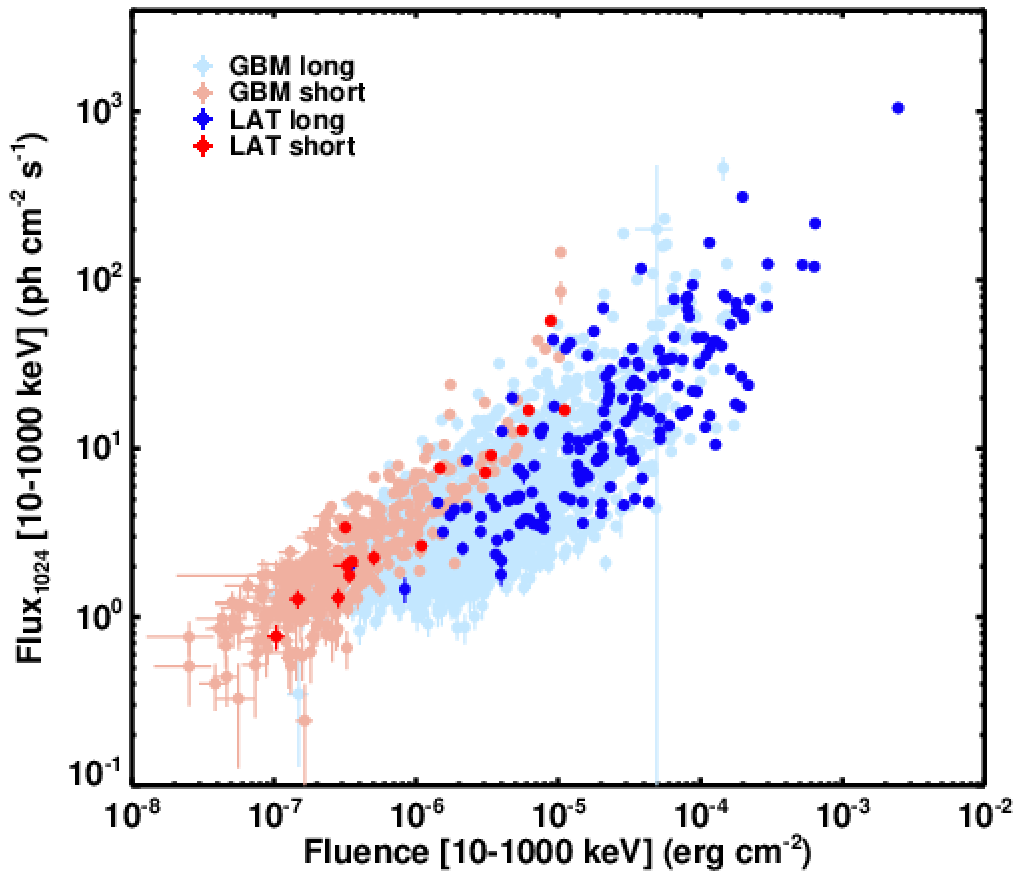} &
\includegraphics[height=0.28\columnwidth]{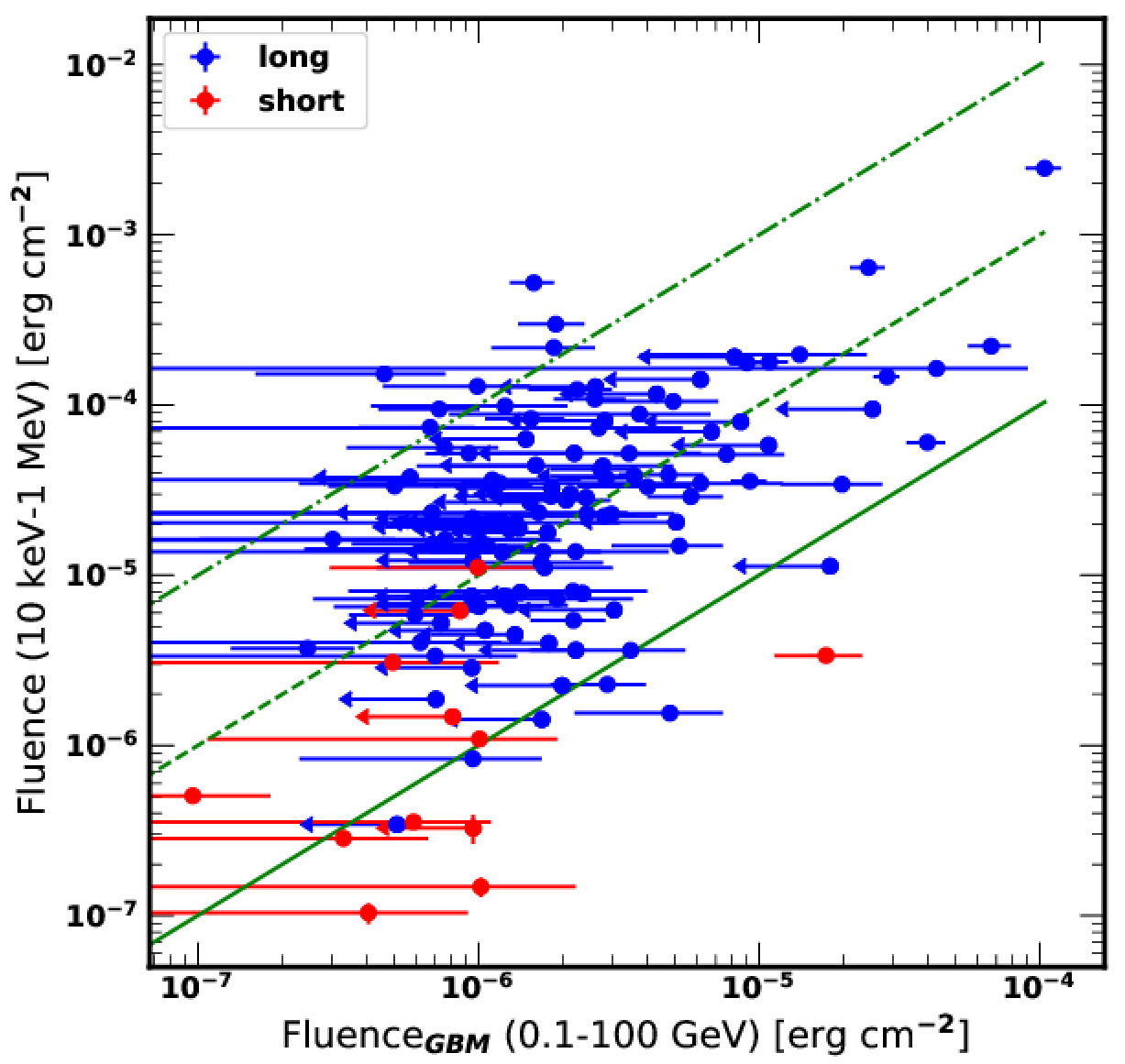} & 
\includegraphics[height=0.28\columnwidth]{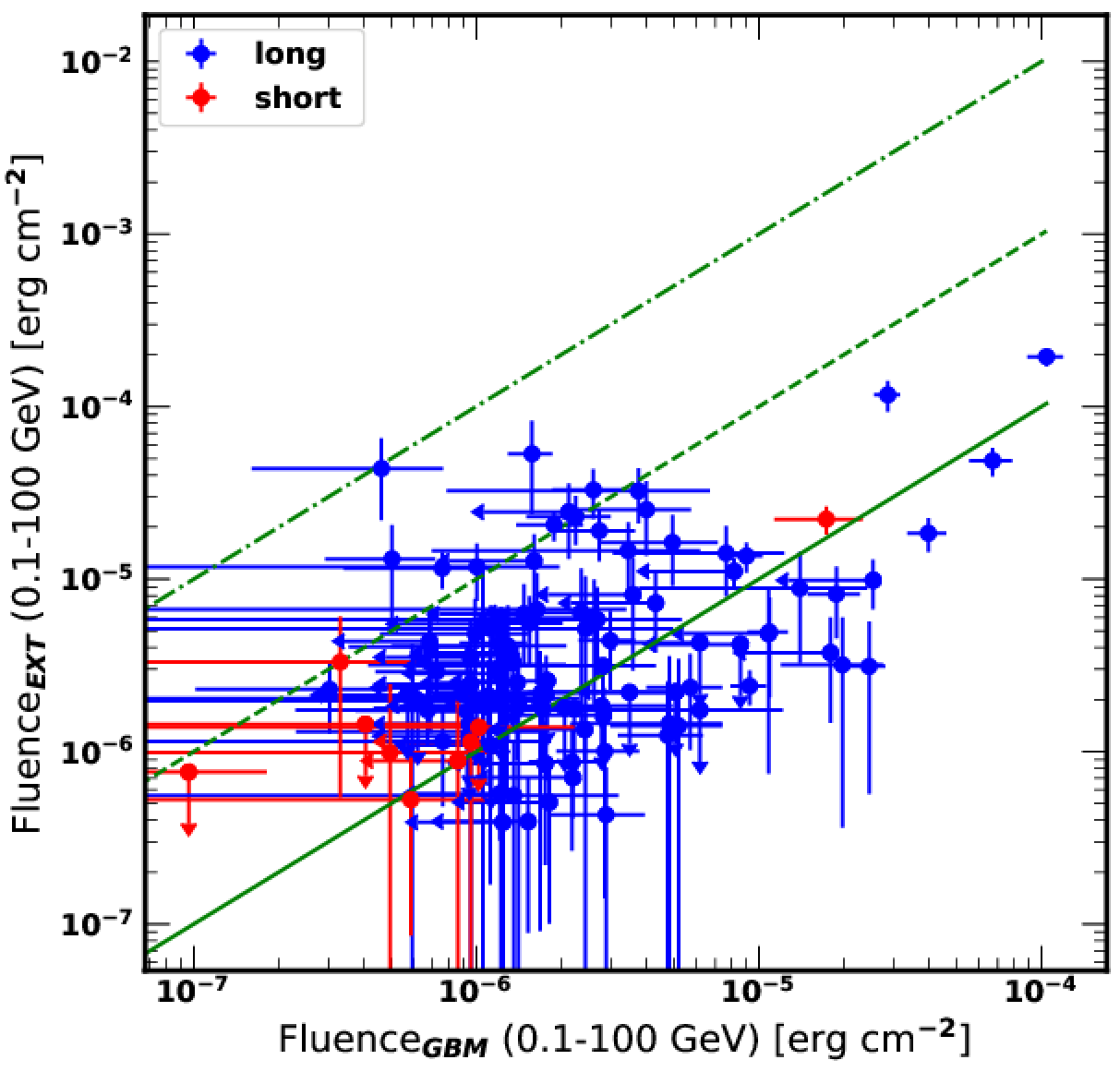}
\end{tabular}
\caption{Left panel: The distribution of energy fluence calculated in the 10--1000\,keV energy range for bursts detected by the LAT compared to the entire sample of 2357 GRBs detected by GBM over the same time period. Middle panel: Fluences calculated in the 10--1000 keV energy range vs. fluences calculated in the 0.1--100 GeV energy range. All values are estimated in the prompt time window as measured by the GBM.  The solid green lines denote where values are equal. The dashed and dashed-dotted green lines are shifted by factors of 10 and 100, respectively.
Right panel: Fluences calculated in the 0.1--100 GeV energy range evaluated in the time window after the GBM emission vs. the same quantities evaluated in the prompt time window. Figure from \cite{2019ApJ...878...52A}.}
\label{fig_fluence}
\end{figure}

Previous studies have indicated that \lat\ preferentially detects GRBs with high fluence at lower energies. Figure~\ref{fig_fluence} compares the values for GRBs included in this study, with the whole sample of GRBs detected by the GBM over the same 10-year period.

The left panel of Figure~\ref{fig_fluence} first compares the fluence in the 10--1000 keV range. While there is a bias for the LAT to detect more fluent bursts, there is a large spread, and also some fainter ones have been detected. This is espceially true for short GRBs (red points). The middle panel compares the fluence at low and high energy and shows that the fluence is dominated by emission at lower energies, which on average is about an order of magnitude larger. The right panel instead compares the high-energy fluence (0.1--100 GeV) in the prompt phase (the GBM \tn) and at later times. Interestingly, the values are often similar, meaning that a significant part of the high-energy flux comes at later times, after the low-energy emission (and presumably the central engine activity) has ceased.
\subsection{High-energy photon index}
The spectral energy distribution above 100 MeV is generally well-fit by a single power law. The photon index shows no sign of being correlated neither with the GBM duration nor the flux (either during prompt or extended times), and is similar for long and short GRBs; however, there is some tendency that the spectra of short-duration GRBs tend to be harder. For both classes of GRBs, the index is compatible with $-2$.
\section{VHE emission from GRBs}
The maximum energy of a photon detected by \lat\ from a GRB so far is 95 GeV from GRB130427A \cite{2014Sci...343...42A}, and only a few GRBs have reached energies above 50 GeV. However, the recent detection of very high-energy (VHE) emission from GRBs by both the H.E.S.S.\footnote{Presented 2019 May 8 at the 1st International CTA symposium} and MAGIC \cite{2019GCN.23701....1M} experiments, show that GRBs are capable of producing emission at much greater energies. We have therefore studied the highest-energy photons for all the GRBs in our study. 
\begin{figure}[t!]
\begin{center}
\includegraphics[width=0.45\columnwidth]{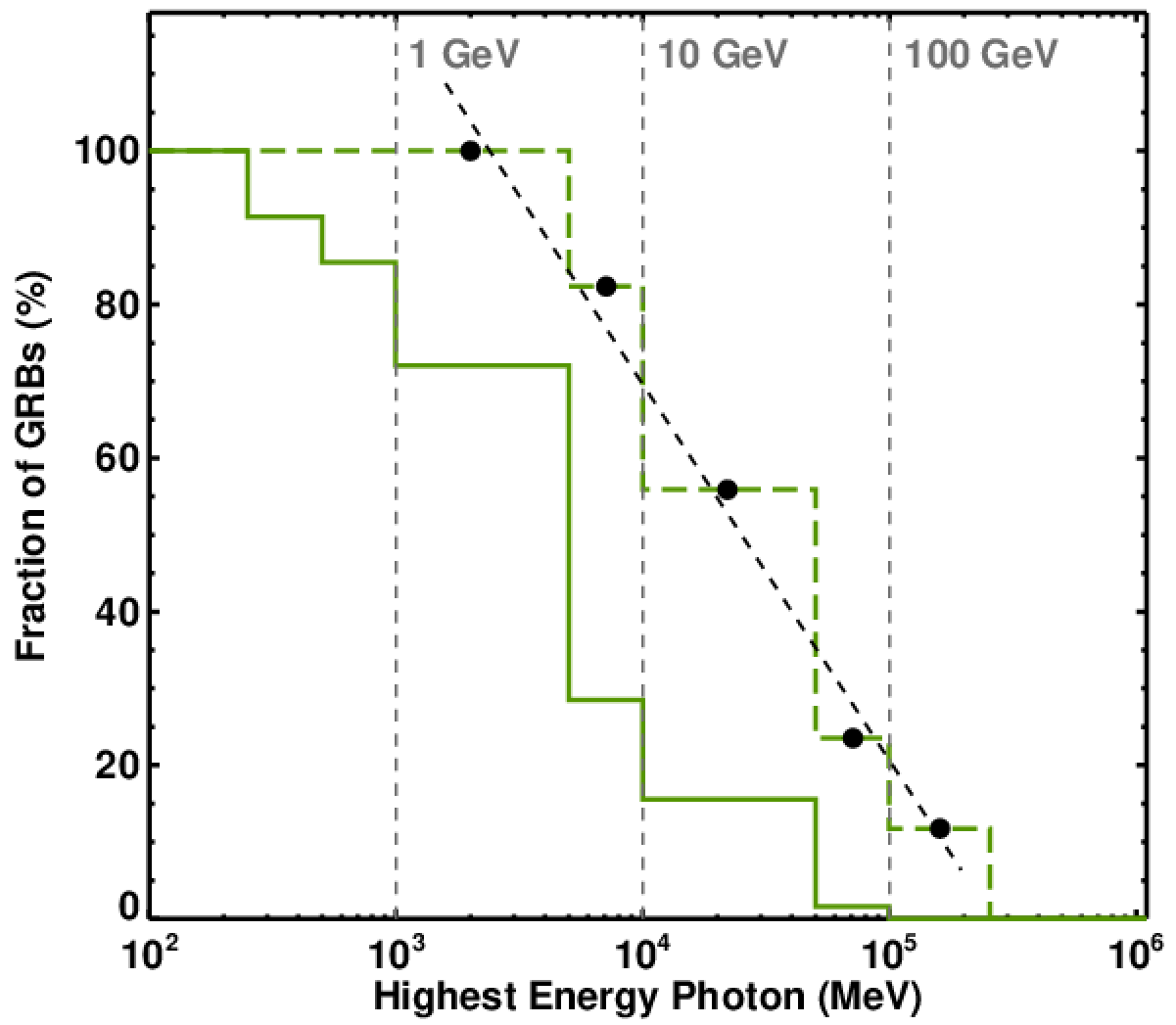}
\caption{Fraction of GRBs with the highest energy photon detected above selected threshold energies (250 MeV, 500 MeV, 1 GeV, 5 GeV, 10 GeV, 50 GeV) ({\it green solid line}). The distribution of the source-frame-corrected energies for the redshift sample is indicated with the {\it dashed green line}. The {\it dashed black line} denotes a linear fit to the values corresponding to the center of each bin. Figure from \cite{2019ApJ...878...52A}.}
\label{HEP_distr}
\end{center}
\end{figure}

Figure~\ref{HEP_distr} shows the fraction of GRBs detected above selected energy thresholds (250 MeV, 500 MeV, 1 GeV, 5 GeV, 10 GeV, 50 GeV). A sharp drop from $\sim$70\,\%\ to $\sim$30\,\%\ is seen at 5 GeV. There are three GRBs with emission above 50 GeV (2\,\%), namely GRB 130427A (95 GeV), GRB 140928A (52 GeV) and GRB 160509A  (52 GeV). 

In our sample, there are also 34 GRBs with measured redshift, which allow us to study the source-frame-corrected energies. This is shown as the dashed line in Figure~\ref{HEP_distr}. The distribution shows a more gradual decrease with energy: almost 80\,\% of the included GRBs have a maximum source-frame photon energy above 5\,GeV, and $\sim$12\,\% (4 GRBs) above 100\,GeV. The highest source-frame energy is a 147\,GeV photon from GRB 080916C, at $z=4.35$ \cite{2013ApJ...774...76A}. 
The figure also includes a linear fit to the bin centers of the source-frame distribution. The fit is remarkably good, showing that the fraction of GRBs decreases as $A\times \log(E/1\,{\rm MeV})+B$, where $A = -49 \pm 4$ and $B = 266 \pm 21$. It is tempting to connect this behavior to the underlying spectral distribution, such that we are seeing the limit determined by the intrinsic spectral shape, which seems to be similar for all GRBs. In bright bursts more high-energy emission is produced, allowing GeV emission to be observed. Faint bursts will produce too little GeV emission and the LAT will only detect MeV photons.

\begin{figure}[t!]
\begin{center}
\begin{tabular}{cc}
\includegraphics[height=0.3\textheight]{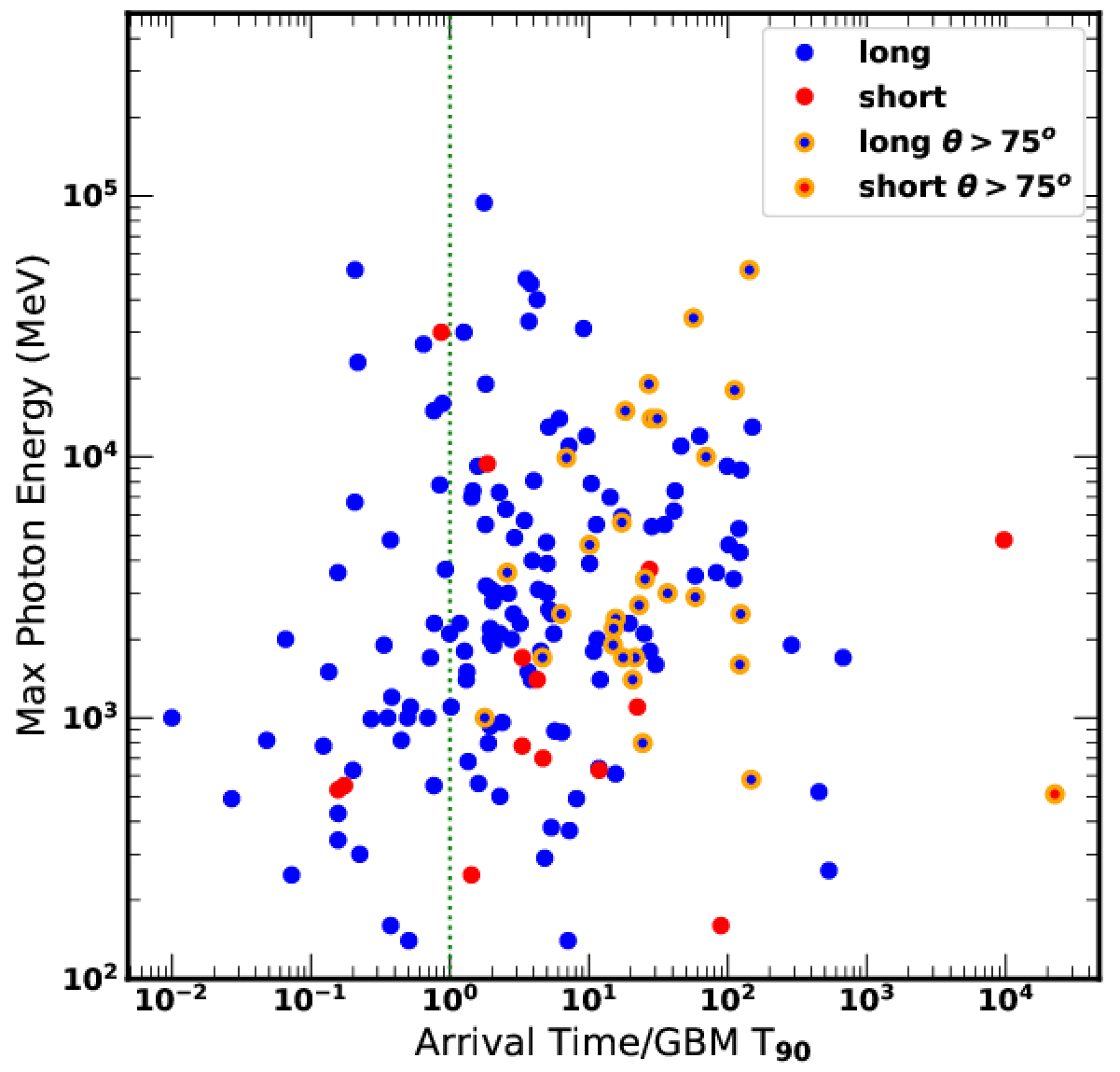} & 
\includegraphics[height=0.3\textheight]{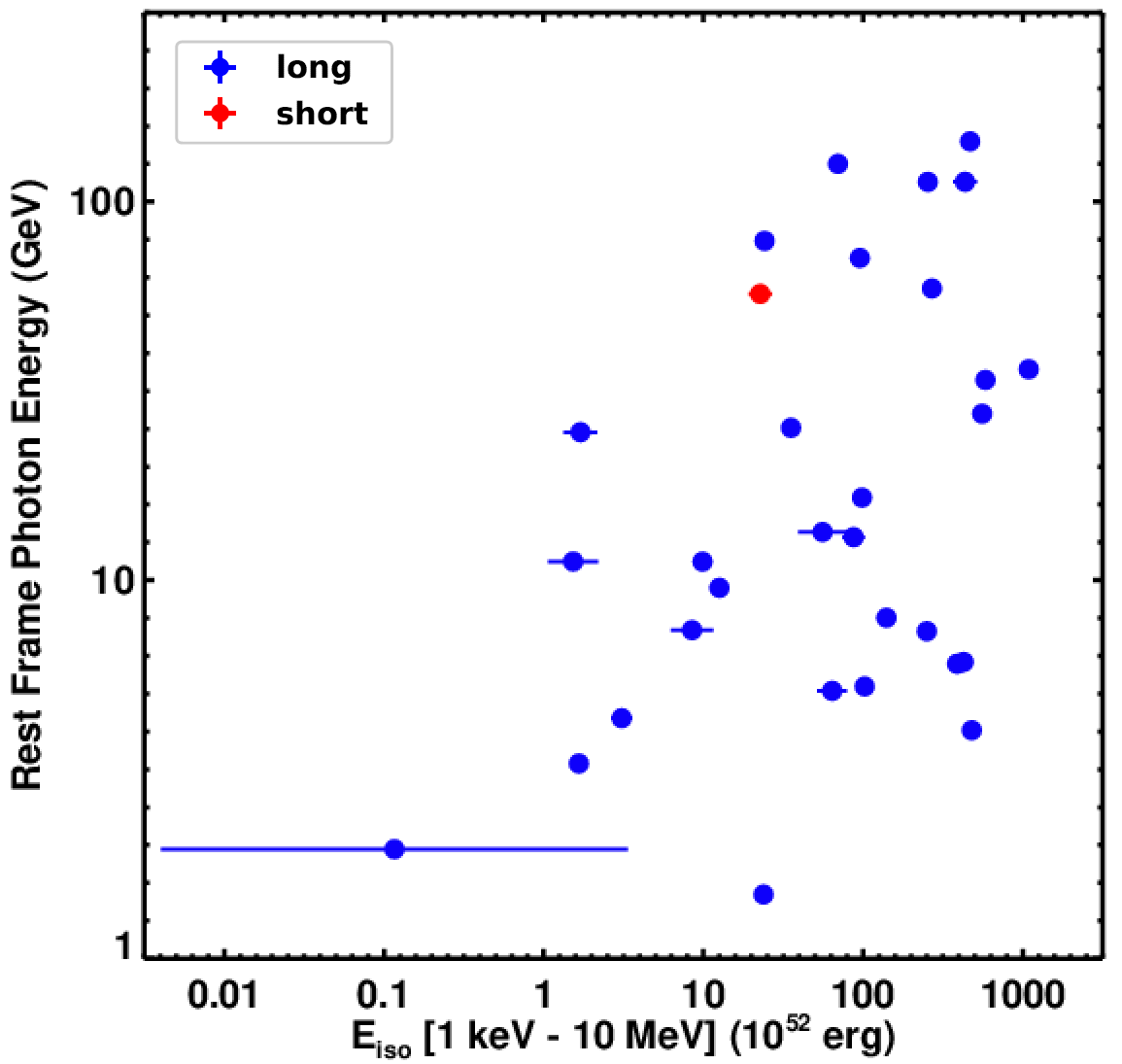}
\end{tabular}
\caption{Left panel: Energy vs. arrival time for the highest energy photon of each GRB. The arrival time is normalized to the duration (T$_{90}$) calculated in the 50--300 keV energy range (indicated by the {\it dashed vertical line}). Right panel: Energy of the highest-energy photon calculated in the rest frame vs. $E_{\rm iso}$ in the 1 keV--10 MeV range. In both panels, {\it blue} and {\it red circles} represent long and short GRBs, respectively. Figure from \cite{2019ApJ...878...52A}.}
\label{HEP_En_Time}
\end{center}
\end{figure}

The left panel of Fig.~\ref{HEP_En_Time} shows the energy of the highest-energy photon in each GRB as a function of arrival time, normalized by the duration in the 50--300\,keV range. No clear pattern can be distinguished, with long and short bursts overlapping in the right panel. The one clear outlier in the right panel is the short GRB 170127C, where the highest energy photon (500\,MeV) was detected almost 3\,ks after the trigger. This GRB was outside the LAT FoV at the trigger time, and the data therefore only cover the time from around 300\,s after the trigger.

Although it is clear that the high-energy emission in general can last a long time, the highest-energy photon in a given GRB in some cases arrives relatively soon after the trigger. For example, this photon arrives within $<2$\,s from the trigger time for $\sim50$\%\ of the sGRBs. Figure~\ref{HEP_En_Time} further shows that also among the GRBs with a maximum photon energy $>10$\,GeV, the highest energy photon is in some cases detected before the emission in the 50--300 keV energy range is over. Particle acceleration in GRBs must be efficient in order to produce such high-energy gamma rays within such a short time. Considering internal opacity of the jet outflow also leads us to conclude that a high bulk Lorentz factor is necessary in order for us to be able to detect these photons.

Previous results have hinted at a trend for the photons with highest source-frame energy to appear in the GRBs with highest isotropic energy ($E_{\rm iso}$) in the 1 keV--10 MeV range. With the greater number of GRBs with known redshift in our study, the trend not only persists but is also extended over a greater range, as shown in the right panel of Figure~\ref{HEP_En_Time}. This indicates that the maximum rest-frame energy produced is simply a function of the total energy output. However, there is a small number of GRBs that deviate from the general trend, showing low rest-frame energies despite a large $E_{\rm iso}$. More detailed studies are required in order to understand this behavior. In both panels of Figure~\ref{HEP_En_Time}, short and long GRBs occupy the same region. This points to similar conditions for the two classes.

\section{Conclusions}
In this contribution we presented a summary of the main results from the second \lat\ GRB catalog. Compared to the first GRB catalog, several changes and improvements to the analysis have been implemented, resulting in a increase from $<12$ to $>18$ GRB detections per year. The results of the analysis for each of the 186 included GRBs are publicly available on the HEASARC web site\footnote{\url{https://heasarc.gsfc.nasa.gov/W3Browse/fermi/fermilgrb.html}}.

The 2FLGC corroborates some of the conclusions from the first catalog, such as the delayed onset and longer duration for the high-energy emission. The larger sample in this catalog is also reflected in the 34 GRBs with measured redshift. This has allowed us to better investigate the energetics also in the rest frame, for example probing the highest-energy photons. The results show that while detections of high-energy photons remain rare, a significant fraction of GRBs have the potential to produce photons well over the 100 GeV limit seen so far. Furthermore, high energies can be achieved fairly quickly after the trigger (as seen in the MAGIC detection of GRB 190114C), but also come much later (as in the H.E.S.S. detection of GRB 180720B). Future detections will thus require a combination of rapidly initiated observations and extended follow-up. With its wide field of view, Fermi will remain a key component for such detections for the foreseeable future.




\acknowledgments
The \lat\ Collaboration acknowledges support for LAT development, operation and data analysis from NASA and DOE (United States), CEA/Irfu and IN2P3/CNRS (France), ASI and INFN (Italy), MEXT, KEK, and JAXA (Japan), and the K.A.~Wallenberg Foundation, the Swedish Research Council and the National Space Board (Sweden). Science analysis support in the operations phase from INAF (Italy) and CNES (France) is also gratefully acknowledged. This work was performed in part under DOE Contract DE-AC02-76SF00515, and has received support from the European Union's Horizon 2020 research and innovation programme under the Marie Sk{\l}odowska-Curie grant agreement No 734303 (NEWS).


\begin{thebibliography}{99}
\bibitem{2009ApJ...697.1071A} W.~B.~Atwood, et al., {\it The Large Area Telescope on the Fermi Gamma-Ray Space Telescope Mission}, ApJ {\bf 697}, 1071 (2009) 
\bibitem{2009ApJ...702..791M}  C.~Meegan, et al., {\it The Fermi Gamma-Ray Burst Monitor},  ApJ {\bf 702}, 791 (2009) 
\bibitem{2019ApJ...878...52A} M.~Ajello, et al., {\it A Decade of Gamma-Ray Bursts Observed by Fermi-LAT: The Second GRB Catalog},  ApJ {\bf 878}, 52 (2019) 
\bibitem{2004ApJ...611.1005G} N.~Gehrels, et al., {\it The Swift Gamma-Ray Burst Mission},  ApJ {\bf 611}, 1005 (2004) 
\bibitem{2013ApJS..209...11A} M.~Ackermann, et al., {\it The First Fermi-LAT Gamma-Ray Burst Catalog},  ApJS {\bf 209}, 11 (2013) 
\bibitem{2002ApJ...579..386N} J.~P.~Norris, {\it Implications of the Lag-Luminosity Relationship for Unified Gamma-Ray Burst Paradigms},  ApJ {\bf 579}, 386 (2002)
\bibitem{2012ApJ...757L..31A} M.~Axelsson, et al., {\it GRB110721A: An Extreme Peak Energy and Signatures of the Photosphere}, ApJ {\bf 757}, 31 (2012) 
\bibitem{PoS(IFS2017)065} E.~Bissaldi, et al., {\it Exploring the low-energy domain of LAT-detected GRBs}, \href{https://pos.sissa.it/312/065/}{PoS (IFS2017) 065}
\bibitem{2018ApJ...864..163V} G.~Vianello, et al., {\it The Bright and the Slow - GRBs 100724B and 160509A with High-energy Cutoffs at $\leq$100 MeV}, ApJ {\bf 864}, 163 (2018) 
\bibitem{2014Sci...343...42A} M.~Ackermann, et al., {\it Fermi-LAT Observations of the Gamma-Ray Burst GRB 130427A}, Science {\bf 343}, 42 (2014) 

\bibitem{2019GCN.23701....1M} R. Mirzoyan et al., {\it MAGIC detects the GRB 190114C in the TeV energy domain}, GRB Coordinates Network 23701, 1 (2019)

\bibitem{2013ApJ...774...76A} W.~B.~Atwood, et al., {\it New Fermi-LAT Event Reconstruction Reveals More High-energy Gamma Rays from Gamma-Ray Bursts}, ApJ {\bf 774}, 76 (2013)

  
%
%
%
\end{thebibliography}
\end{document}